\documentclass[a4paper]{jpconf}
\usepackage{graphicx}
\usepackage{amsmath}
\begin{document}
\title{MRPC Telescope Simulation for the Extreme Energy Events Experiment}

\author{G.~Mandaglio$^{ab}$,  M.~Abbrescia$^{ac}$,
C.~Avanzini$^{ad}$, L.~Baldini$^{ad}$, R.~Baldini~Ferroli$^{ae}$,
G.~Batignani$^{ad}$, M.~Battaglieri$^{af}$, S.~Boi$^{ag}$,
E.~Bossini$^{ah}$, F.~Carnesecchi$^{ai}$, C.~Cical\`{o}$^{ag}$,
L.~Cifarelli$^{ai}$, F.~Coccetti$^{a}$, E.~Coccia$^{aj}$,
A.~Corvaglia$^{ak}$, D.~De~Gruttola$^{al}$, S.~De~Pasquale$^{al}$,
F.~Fabbri$^{ae}$, A. Fulci$^b$, L.~Galante$^{am}$, P.~Galeotti$^{am}$,
M.~Garbini$^{ai}$, G.~Gemme$^{af}$, I.~Gnesi$^{am}$,
S.~Grazzi$^{a}$, D.~Hatzifotiadou$^{ain}$, P.~La~Rocca$^{ao}$, Z.~Liu$^{anp}$,
 G.~Maron$^{q}$, M.~N.~Mazziotta$^{ar}$,
A.~Mulliri$^{ag}$, R.~Nania$^{ai}$, F.~Noferini$^{ai}$,
F.~Nozzoli$^{as}$, F.~Palmonari$^{ai}$, M.~Panareo$^{ak}$,
M.~P.~Panetta$^{ak}$, R.~Paoletti$^{ah}$, C.~Pellegrino$^{ai}$,
L.~Perasso$^{af}$, C.~Pinto$^{ao}$, G.~Piragino$^{am}$,
S.~Pisano$^{ae}$, F.~Riggi$^{ao}$, G.~Righini$^{a}$,
C.~Ripoli$^{al}$, M.~Rizzi$^{ac}$, G.~Sartorelli$^{ai}$,
E.~Scapparone$^{ai}$, M.~Schioppa$^{at}$, A.~Scribano$^{ad}$,
M.~Selvi$^{ai}$, G.~Serri$^{ag}$, S~ Squarcia$^{af}$,
M.~Taiuti$^{af}$, G.~Terreni$^{ad}$, A.~Trifir\`{o}$^{ab}$,
M.~Trimarchi$^{ab}$, A.S. Triolo$^b$, C.~Vistoli$^{q}$, L.~Votano$^{au}$,
M.~C.~S.~Williams$^{ain}$, A.~Zichichi$^{ain}$, R.~Zuyeuski$^{an}$}

\address{

$^a$Museo Storico della Fisica e Centro Studi e Ricerche Enrico
Fermi, Roma, Italy

$^b$INFN Sezione di Catania and Dipartimento di Scienze Matematiche
e Informatiche, Scienze Fisiche e Scienze della Terra,
Universit\`{a} di Messina, Messina, Italy

$^c$INFN and Dipartimento Interateneo di Fisica, Universit\`{a} di
Bari, Bari, Italy

$^d$INFN and Dipartimento di Fisica, Universit\`{a} di Pisa, Pisa,
Italy

$^e$INFN, Laboratori Nazionali di Frascati, Frascati (RM), Italy

$^f$INFN and Dipartimento di Fisica, Universit\`{a} di Genova,
Genova, Italy

$^g$INFN and Dipartimento di Fisica, Universit\`{a} di Cagliari,
Cagliari, Italy

$^h$INFN and Dipartimento di Fisica, Universit\`{a} di Siena, Siena,
Italy

$^i$INFN and Dipartimento di Fisica, Universit\`{a} di Bologna,
Bologna, Italy

$^j$INFN and Dipartimento di Fisica, Universit\`{a} di Roma Tor
Vergata, Roma, Italy

$^k$INFN and Dipartimento di Matematica e Fisica, Universit\`{a} del
Salento, Lecce, Italy

$^l$INFN and Dipartimento di Fisica, Universit\`{a} di Salerno,
Salerno, Italy

$^m$INFN and Dipartimento di Fisica, Universit\`{a} di Torino,
Torino, Italy

$^n$CERN, Geneva, Switzerland

$^o$INFN and Dipartimento di Fisica e Astronomia "E.Majorana",
Universit\`{a} di Catania, Catania, Italy

$^p$ICSC World laboratory, Geneva, Switzerland

$^q$INFN-CNAF, Bologna, Italy

$^r$INFN Sezione di Bari, Bari, Italy

$^s$Trento Institute for Fundamental Physics and Applications,
Trento, Italy

$^t$INFN and Dipartimento di Fisica, Universit\`{a} della Calabria,
Cosenza, Italy

$^u$INFN, Laboratori Nazionali del Gran Sasso, Assergi (AQ), Italy

}

\ead{gmandaglio@unime.it}

\begin{abstract}
A simulation tool based on GEMC framework to describe the MRPC telescope of the Extreme Energy Events (EEE) Project is presented. The EEE experiment   is mainly devoted to the study of the
secondary cosmic muons  by using  MRPC telescope distributed in high schools and research centres in Italy and at CERN.  This takes
into account the muon interactions with EEE telescopes and the structures surrounding
the experimental apparata; it consists of a dedicated event generator producing
realistic muon distribution and a detailed geometry description of the detector. Microscopic behaviour of MRPCs has been included to produce experimental-like data. A method to estimate the chamber efficiency directly from data has been implemented and tested by comparing the experimental and simulated polar angle distribution of muons.
\end{abstract}

\section{Introduction}

The EEE experiment\cite{Ref1} is a  \emph{``Museo Storico della Fisica e Centro Studi e
Ricerche Enrico Fermi''}\cite{Ref2} project carried on in collaboration with  \textit{``Istituto Nazionale di Fisica Nucleare''} (INFN) \cite{Ref3},  and \emph{"Ministero dell'Universit\`{a},
dell'Istruzione e della Ricerca"} (MIUR) and CERN\cite{Ref4}.
The experiment consists of a network of 59 MRPC-based telescopes, distributed in the Italian High Schools,  at CERN (two) and in five research centres of INFN, covering an area of about 0.3$\times$10$^6$ km$^2$.
The EEE activities are devoted to both educational and scientific purposes. The experiment has already collected 
more than 100 billion  muon tracks
offering a large scientific programme:  the investigation of extensive air showers by looking for coincidences between different telescopes\cite{riggilong,Ref14},  the investigation of Forbush decrease\cite{forby} and the  monitoring of the building stability by combining the EEE telescope with other detectors\cite{riggistab} etc.
To fully understand the data coming from  telescopes installed in different places,  a precise knowledge of the effect on the measurements of the different structures holding the detectors is needed; the different configuration, such as the distance between the MRPC chambers of a single telescope, must also be taken into account.
For these reasons, we implemented a simulation tool, by using the GEMC (GEant4 MonteCarlo) framework, in order to be able to describe the behaviour of a single telescope working in different places and with different configuration setups, to estimate angular and absolute efficiency, and the absolute single muon rates.

\section{The EEE Detectors and the Simulation Tool}

\begin{figure}[h!]
\centering
\includegraphics[width=0.28\columnwidth]{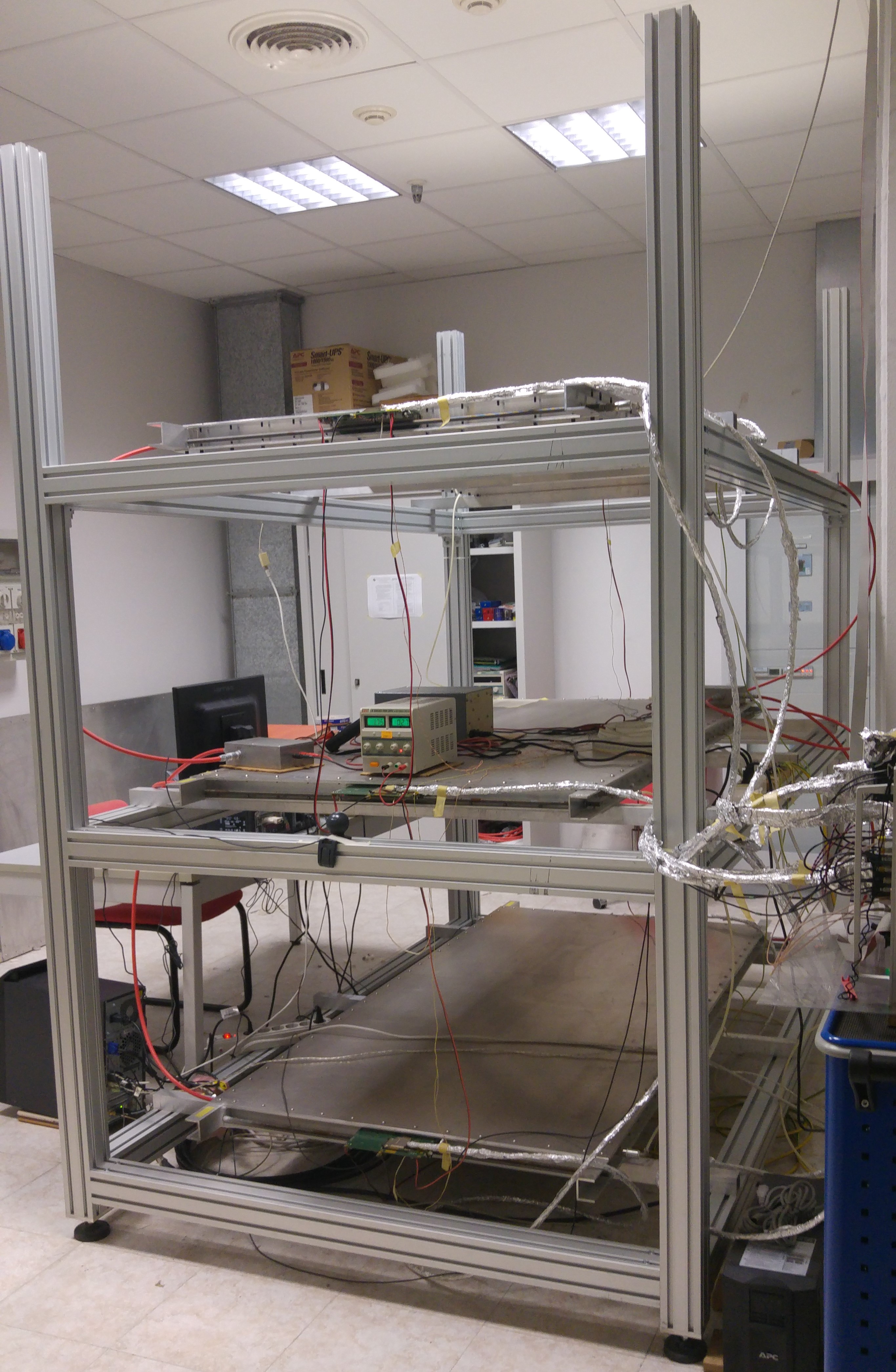}
\caption{Telescope operating at \textit{``INFN - Sezione di Catania''} (Italy).} \label{cata}
\end{figure}

Each EEE telescope consists of three Multigap Resistive Plate Chambers (MRPCs)
with a 80$\times$160 cm$^2$ 
active area, 
assembled in a stack with 50 cm distance between two MRPCs in the most common configuration (see picture of a telescope installed at \textit{``INFN - Sezione di Catania''}, reported in figure \ref{cata}).
Each chamber is segmented by  24 copper strips (180 cm $\times$ 2.5 cm spaced by 7 mm), 
which  collect the charge signals produced in the gas (mixture of C$_2$F$_4$H$_2$ (98\%) and SF$_6$ (2\%)) of the chamber by the crossing of charged particles.
The chamber configuration provides us two-dimensional information:  one is given by the coordinate of the strip giving signal or, in the case of contiguous strips by averaging their positions, while the other one is obtained by 
the time difference of the signals at the opposite edges of the strip (measured using  TDCs with 100ps resolution).
Detailed description of the detector is reported in the paper by P. La Rocca et al. published in the same review volume as the present paper~\cite{paola_desyt}.

\subsection{Event Generator}
 The model described in Ref.\cite{egpar} has been used to generate the single-muon distribution.
The azimuthal distribution of muons is considered to be uniform, while  the polar angle distribution (angle with respect to the Zenith)) is taken from an improved Gaisser-like parametrization (see Ref.~\cite{egpar}) that includes Earth curvature at all latitudes and low energy muons ($E_\mu<100$ GeV/c). The flux, as a function of muon energy $E_\mu$ and $\theta_\mu$ is given by:

\begin{equation}
\begin{aligned}
        \frac{dI_\mu}{dE\mu}=0.14
    \left[ \frac{E_\mu}{GeV}(1+\frac{3.64\,GeV}{E_\mu(cos(\theta^*)^{1.29}} \right]^{-2.7}\\
    \left[ \frac{1}{1+ \frac{1.1\, E_\mu\,cos \theta^*}{115 GeV}} \right]+
    \left[ \frac{0.054}{1+ \frac{1.1\, E_\mu\,cos \theta^*}{850 GeV}} \right]
\end{aligned}
\end{equation}
where 
\begin{equation}
cos\theta^* = 0.14 \sqrt{\frac{(cos\theta)^2+ P^2_1+P_2(cos\theta)^{P_3}+P_4(cos\theta)^{P_5}}{1+P_1^2+P_2+P_4}}
\label{costheta}
\end{equation}
and the used parameters $P_1-P_5$ are 0.102573, -0.068287, 0.958633, 0.040725, 0.817285, respectively.
The parametrization is able to reproduce well the existing measurements, see Fig. 2 of Ref. \cite{egpar}.  
The absolute muon flux normalization of 1.06 cm$^{-2}$ min$^{-1}$, used in the simulation, is the one reported in the PDG \cite{pdg}.
 Finally, the generated muons were processed by GEMC to record the simulated interaction with the EEE telescope and  to derive the expected angular distribution of single-track events  and the absolute rate.

\subsection{GEMC (GEANT4-library-based interface)}

\begin{figure}[h]
\centering
\includegraphics[width=0.38\columnwidth]{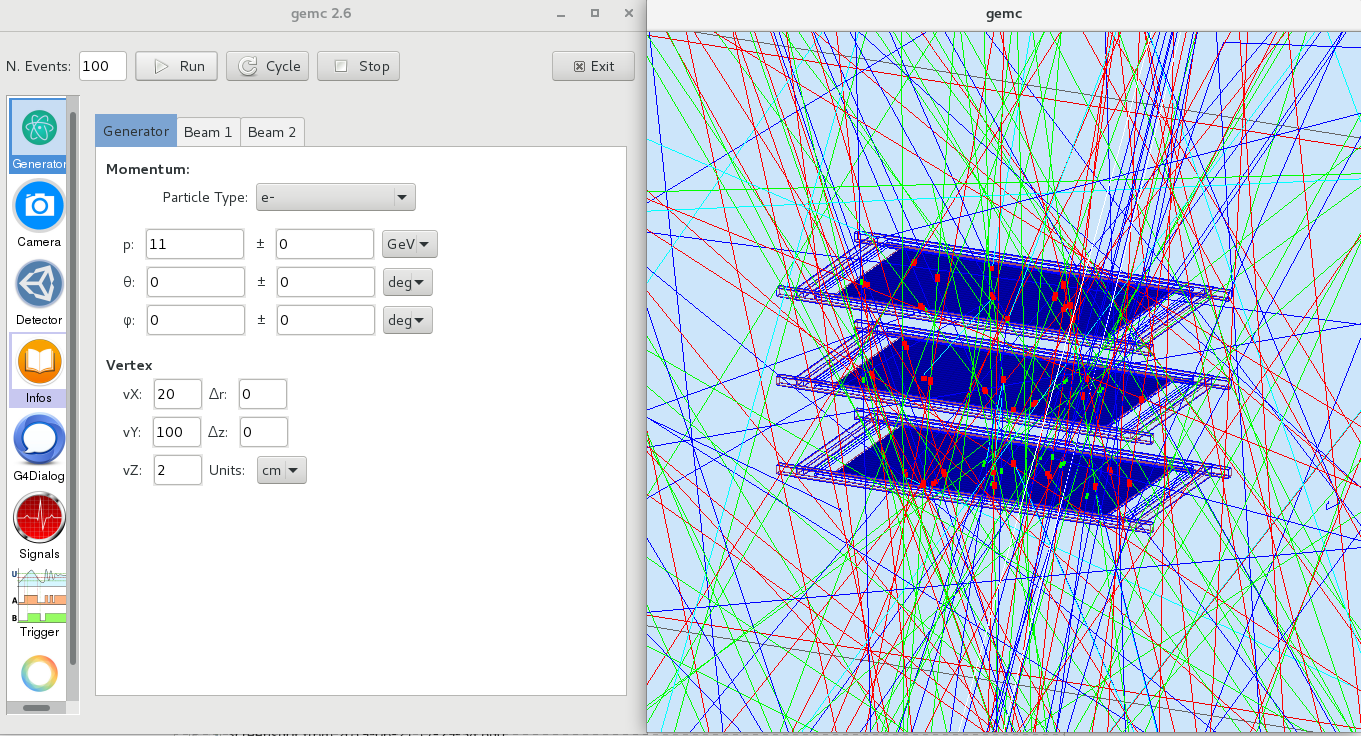}
\includegraphics[width=0.25\columnwidth]{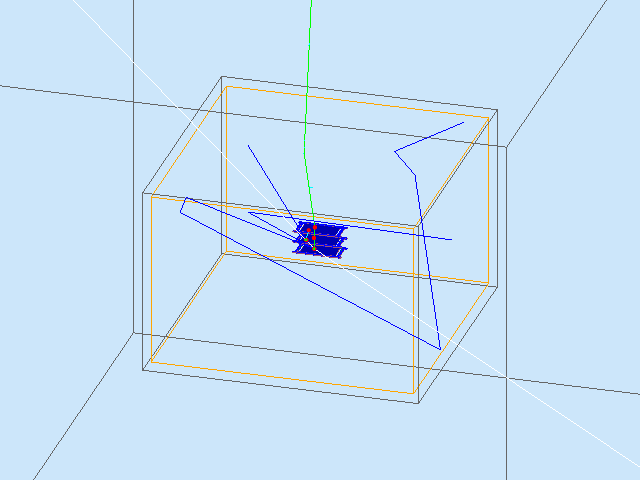}
\includegraphics[width=0.25\columnwidth]{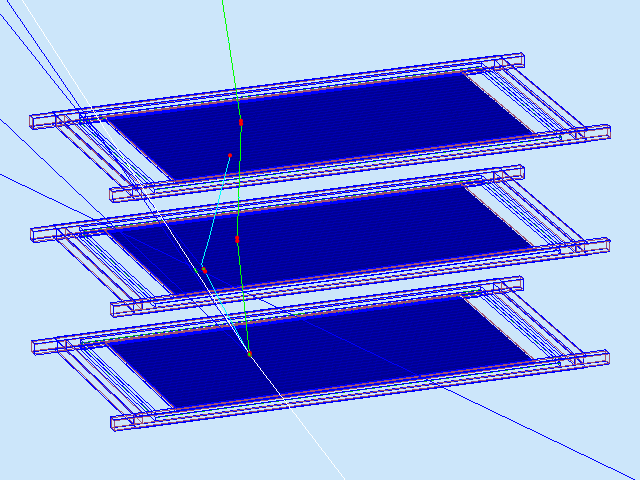}
\caption{GEMC graphical interface (left panel); details of a muon interaction  with the structure of the room (box of 30 cm thick  concrete) hosting the detector (center panel) and a zoom on the telescope (right panel).} \label{geo}
\end{figure}

This simulation tool is based on  the GEMC \cite{gemc}   framework providing  user-defined geometry and hit description. 
Detector and building structures are implemented by using  the standard GEANT volume description. The program handles multiple input/output format and provides a graphical interface to visualize the detector and the hits in active and passive volumes (see figure \ref{geo}).

The MRPC response was parametrized based upon the measured performance of the chambers \cite{degruttola}. In particular, the algorithm that mimics the avalanche propagation in the gas is effectively described by a cone with the vertex generated by the particle interaction in the upper layer of the chamber and that develops downwards to the bottom one. The size of the cone in the transverse plane (with respect to the track propagation) is assumed to be the cluster size  $\sigma_X$=9.2 mm and $\sigma_Y$= 15 mm as measured in Ref. \cite{degruttola}. To reproduce the measured speed and time resolution, the signal propagation time along the strip is assumed to be 15.8  ns/cm,   smeared to take into account the measured experimental resolution of  $\sigma_T=238$ ps. 
In the simulation, the three chambers in the most common configuration are placed in a concrete wall box with variable thickness mimicking the room where the real telescope is located. Of course, more complicated geometries of the surrounding space can be implemented.

The information generated by GEMC and necessary to reconstruct the muon track is: the total number of hits for each chamber (at least one);  the coordinates of the strips that give signals;  the signal time from the generation point to the edge of the chamber. By using this information the reconstruction program is able to write data in the experimental format.

\section{Efficiency Estimation and Experimental-Simulated Data Comparison}

In order to compare  the simulated and experimental data, we calculate the efficiency of a telescope of the network (TORI-03), selected for its stable working condition. We mapped each chamber by dividing it in  24$\times$20 (X$\times$Y directions) sectors, and then we estimated for each bidimensional interval  the tracking efficiency and the counting efficiency, assuming no correlation between the two quantities. 
\begin{figure}[h]
\centering
\includegraphics[width=0.5\columnwidth]{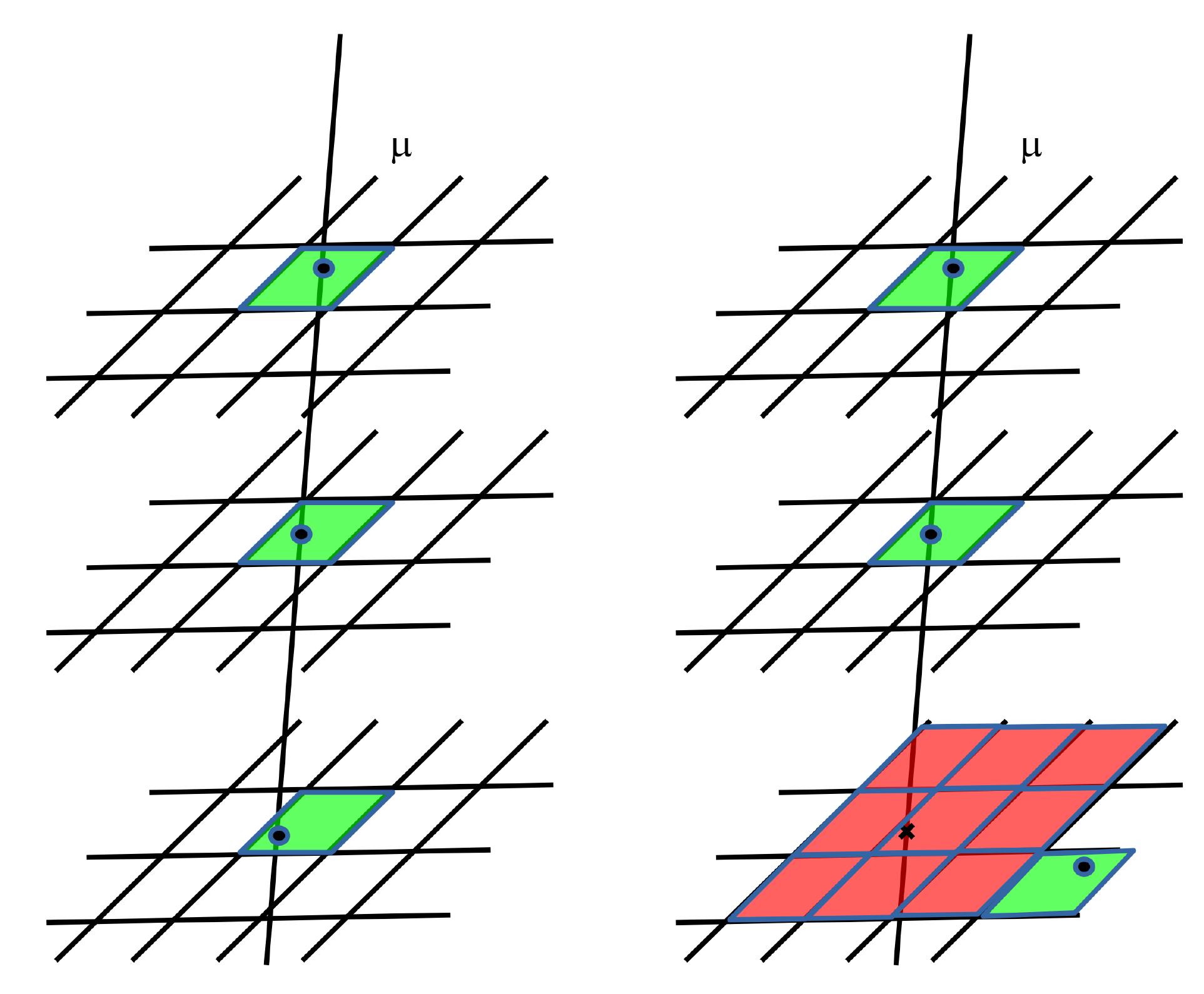}
\caption{Scheme of good coincidence with all hits in the interval area of chamber intercepted by the muon track (left) and the case of a missing hit (right); in green  the areas  having a detected hit.} \label{scheme}
\end{figure} 

We define the tracking efficiency as the ratio between the map of the missing hits (geometrical position belonging to a sector of a chamber crossed by a track reconstructed by using the information provided by the other two chambers) and the map of the good hits (hits belonging to sectors of chambers crossed by tracks reconstructed by using information of all chambers). The cases of missing hit and the good hit are showed in the scheme in figure \ref{scheme}.

\begin{figure}[h]
\centering
\includegraphics[width=0.4\columnwidth]{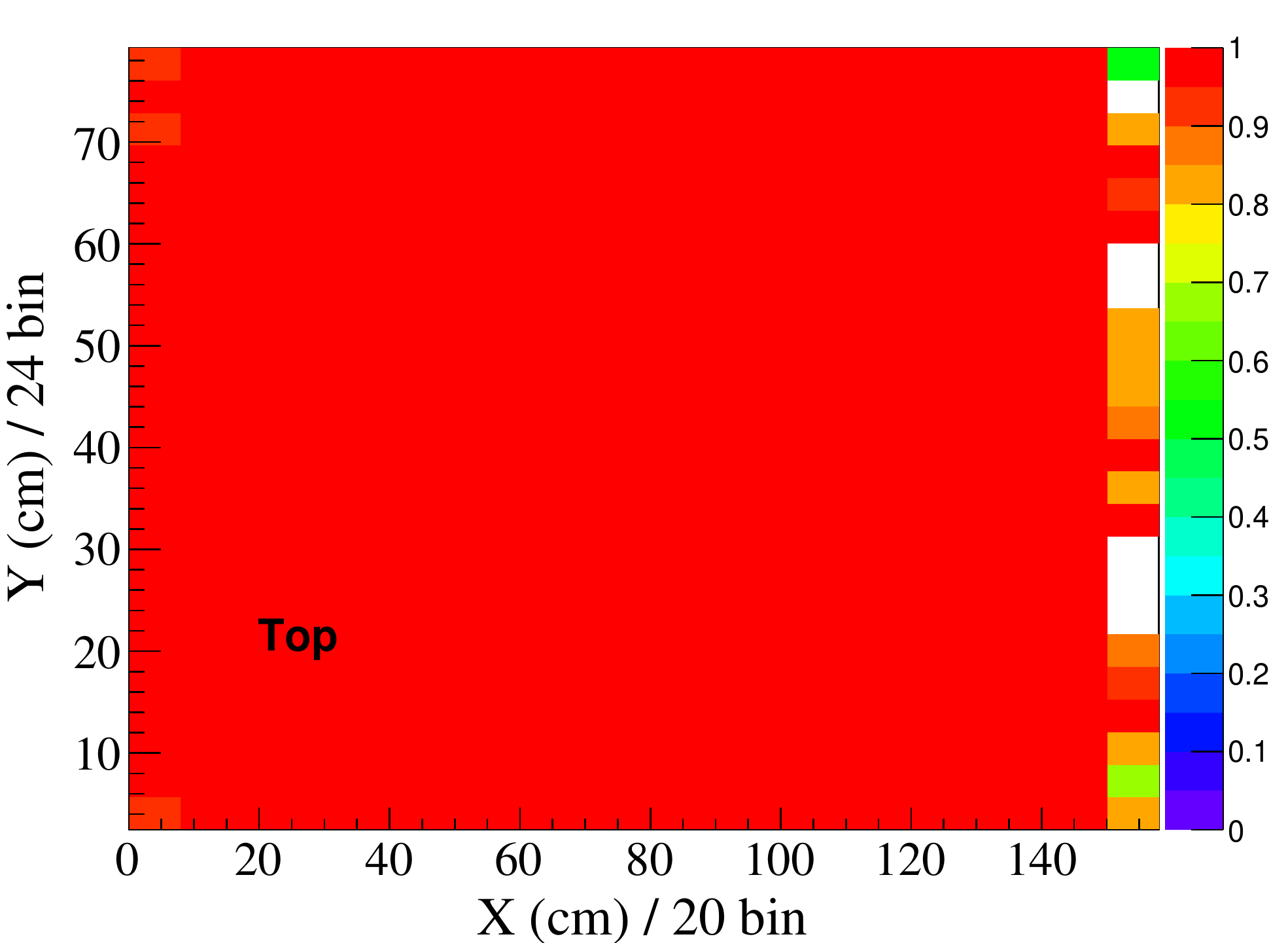}
\put(-100,50){Preliminary}
\includegraphics[width=0.35\columnwidth]{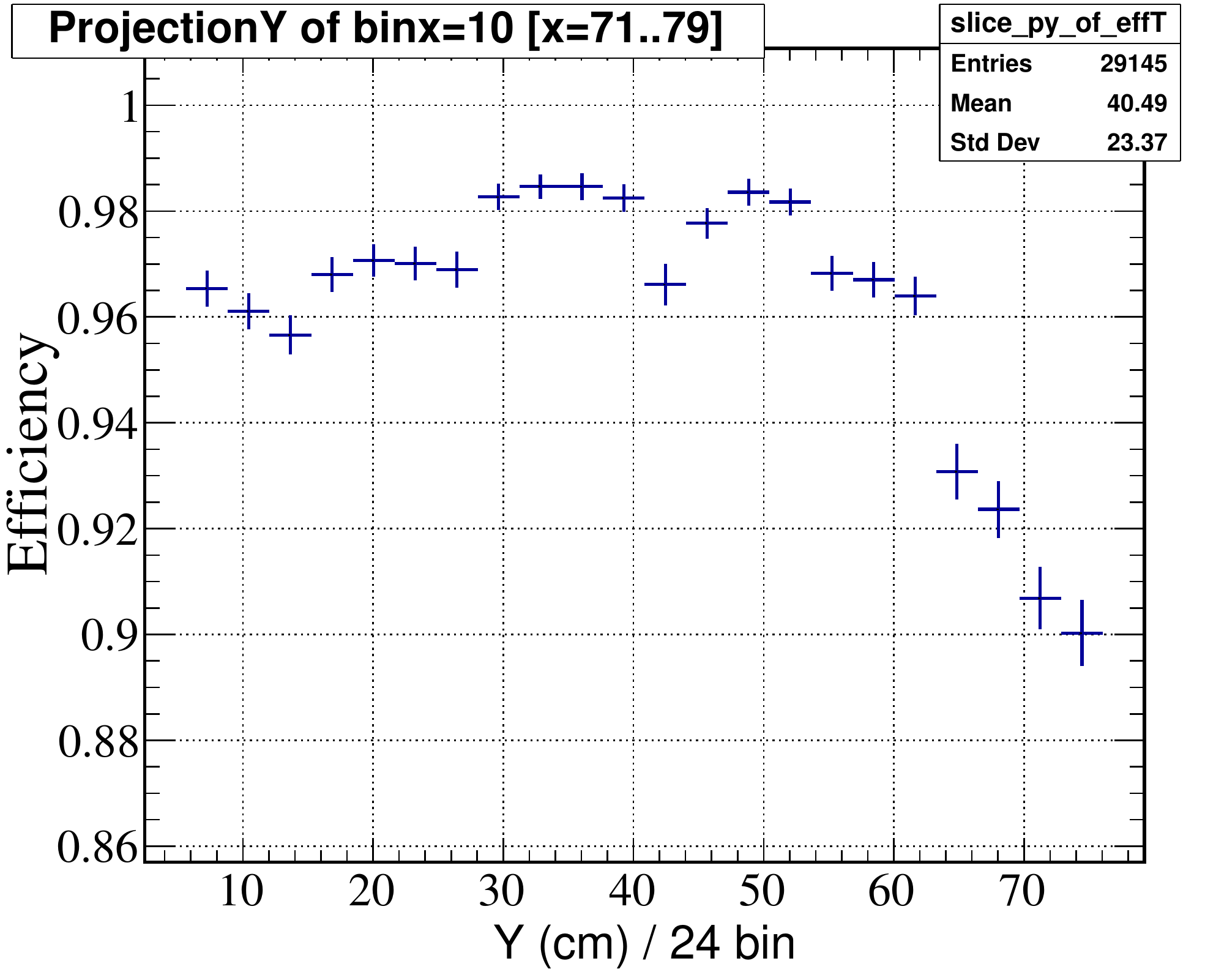}
\put(-100,50){Preliminary}
\caption{Tracking efficiency map of top chamber (left) and projection of the map with respect to the X axis (right) for TORI-03 telescope.} \label{trackeff}
\end{figure}

The tracking efficiency for the top chamber of TORI-03 telescope is reported as an example in figure \ref{trackeff}. By looking at the projection of the efficiency map reported in the right panel of figure \ref{trackeff}, the used algorithm estimates an efficiency higher than 90\% and close to 98\% in the central region of the chamber in the analysed data sample.

\begin{figure}[h]
\centering
\includegraphics[width=0.4\columnwidth]{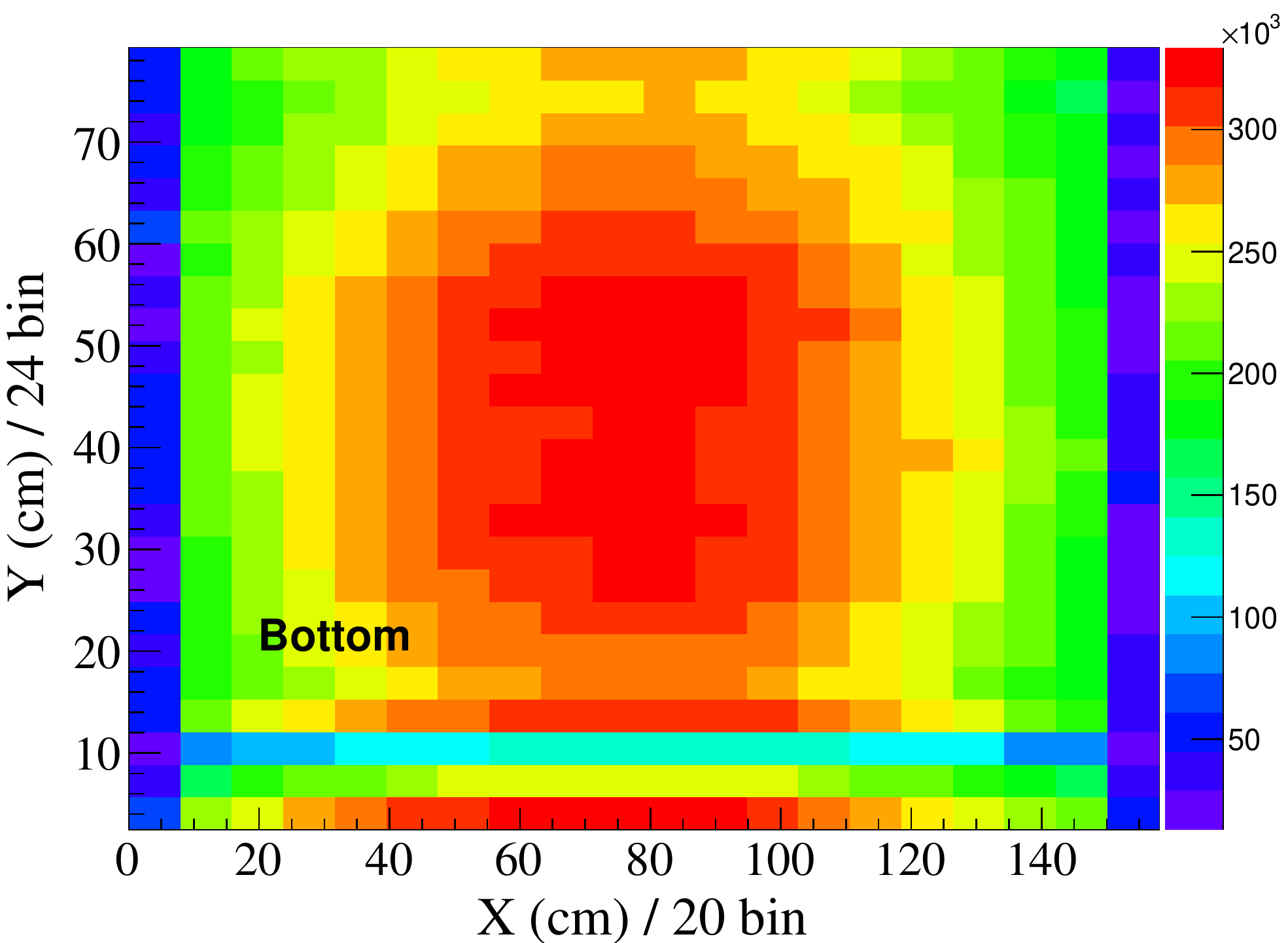}
\put(-100,110){Preliminary}
\includegraphics[width=0.4\columnwidth]{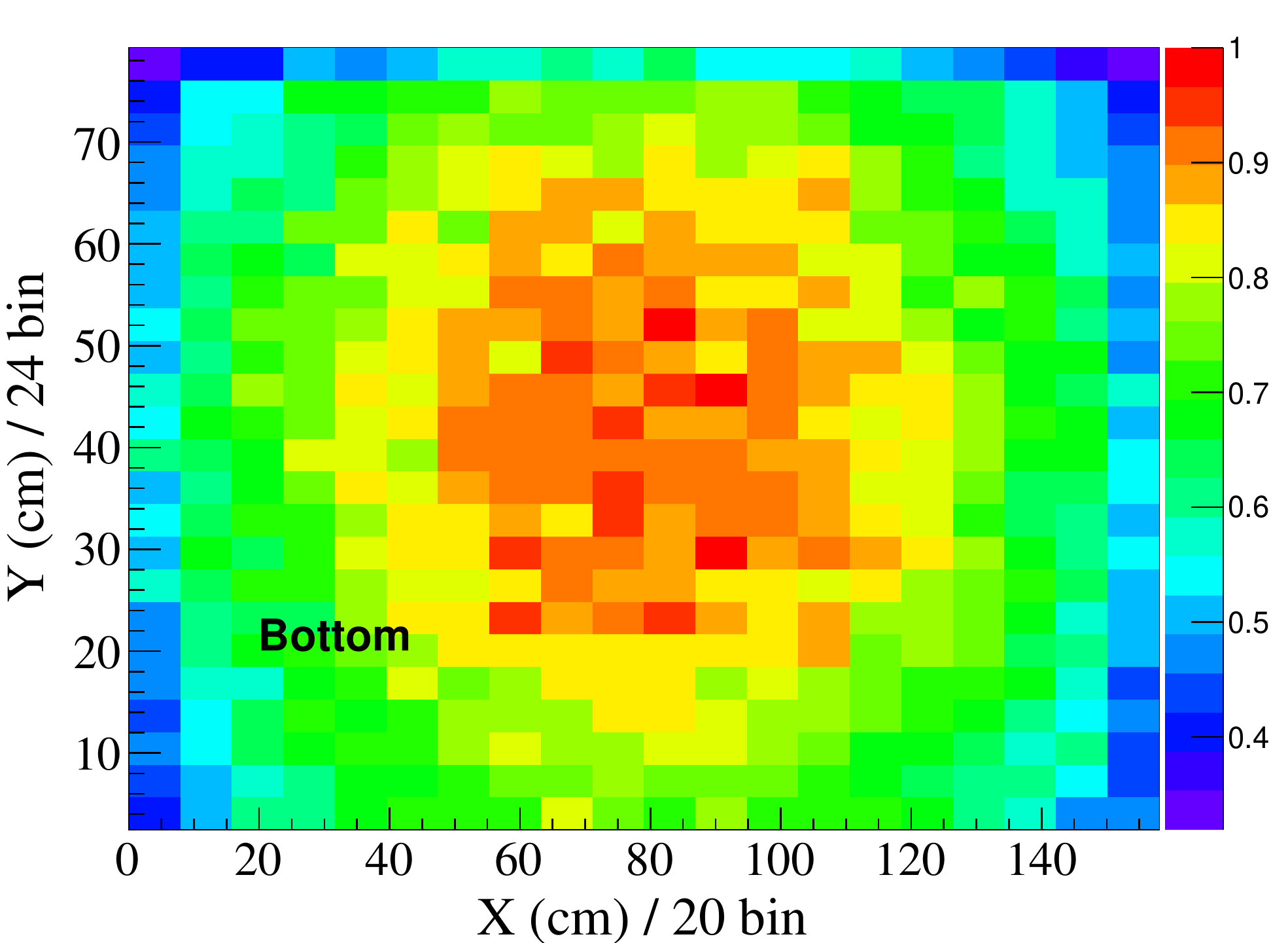}
\put(-100,110){Preliminary}\\
\includegraphics[width=0.4\columnwidth]{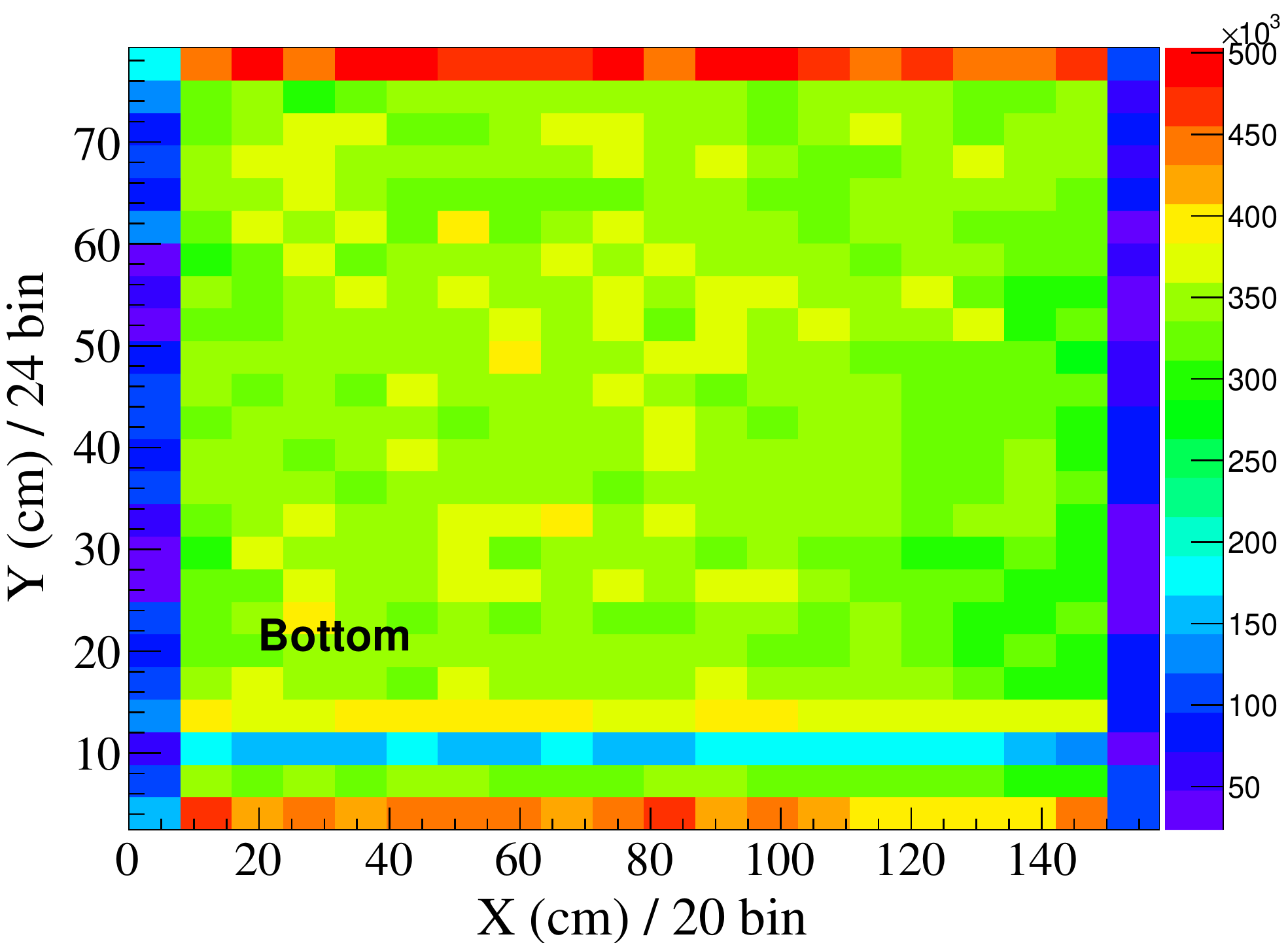}
\put(-100,110){Preliminary}
\includegraphics[width=0.4\columnwidth]{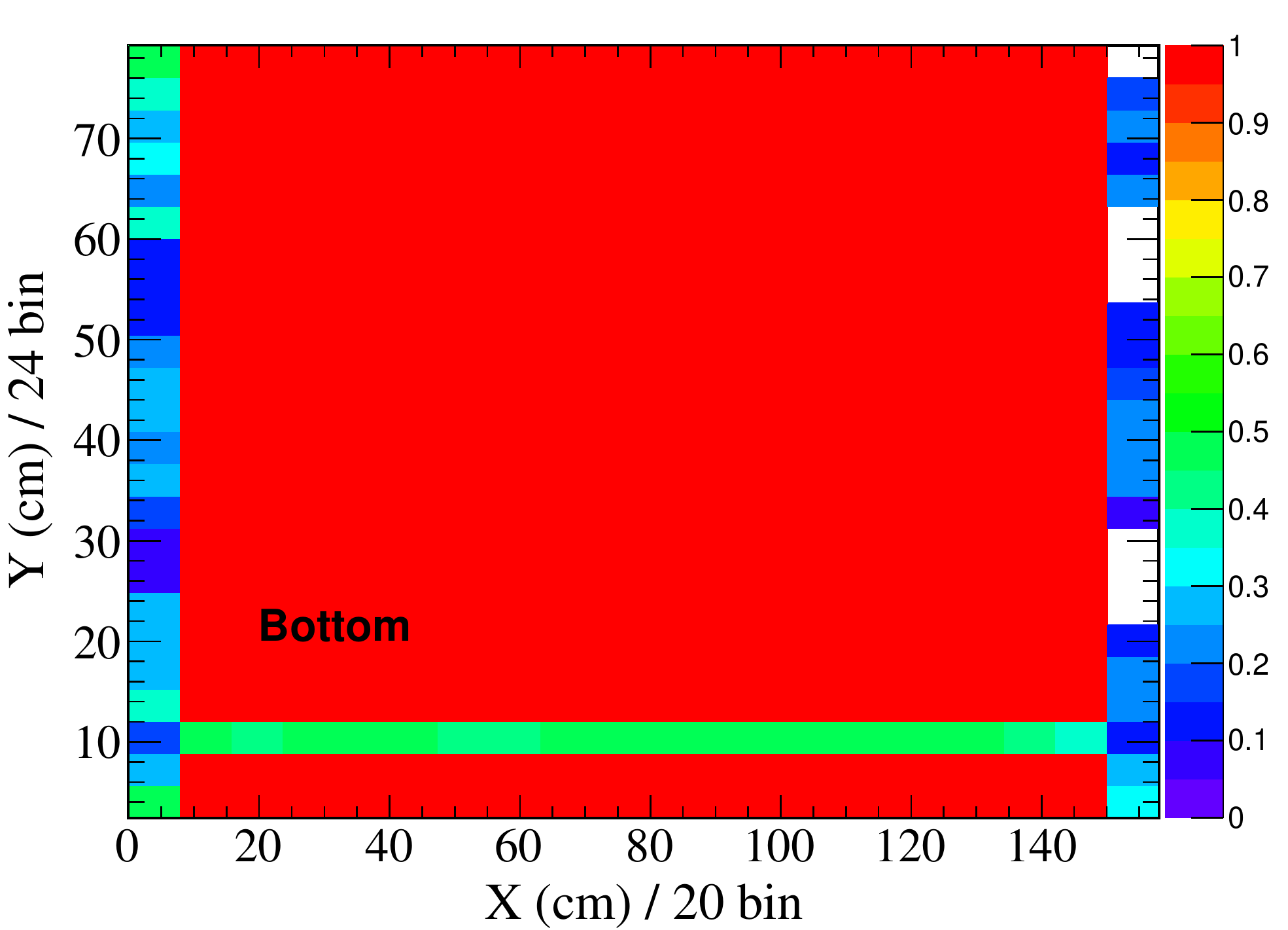}
\put(-100,110){Preliminary}
\caption{Experimental hit map of bottom chamber (top-left) of TORI-03 telescope, simulated hit map normalized to the maximum rate (top-right); the ratio between the top-left and top-right maps (bottom-left); the bottom-right map normalized to the average of rate of the bins in the center of the chamber (bottom-right).} \label{hitrate}
\end{figure}

The counting efficiency is obtained according to the following procedure:\\
1) we compute the map of hit without any constraints (see Fig. \ref{hitrate} top-left);\\
2) we estimate the hit distribution of simulated events (no efficiency information used in the simulation) normalized to the maximum bin rate (see Fig. \ref{hitrate} top-right);\\
3) we remove the effect on hit distribution due to detector acceptance by dividing the map of hit distribution 1) to the map obtained as described in 2), obtaining a flat rate distribution (see Fig. \ref{hitrate} bottom-left);\\
4) finally, we normalize the map resulting from procedure 3) to the average of bins with relative maximum rates computed in the center of chamber to avoid unwanted unstable behaviour at the edges of the detector (bottom-left).

\begin{figure}[h]
\centering
\includegraphics[width=0.33\columnwidth]{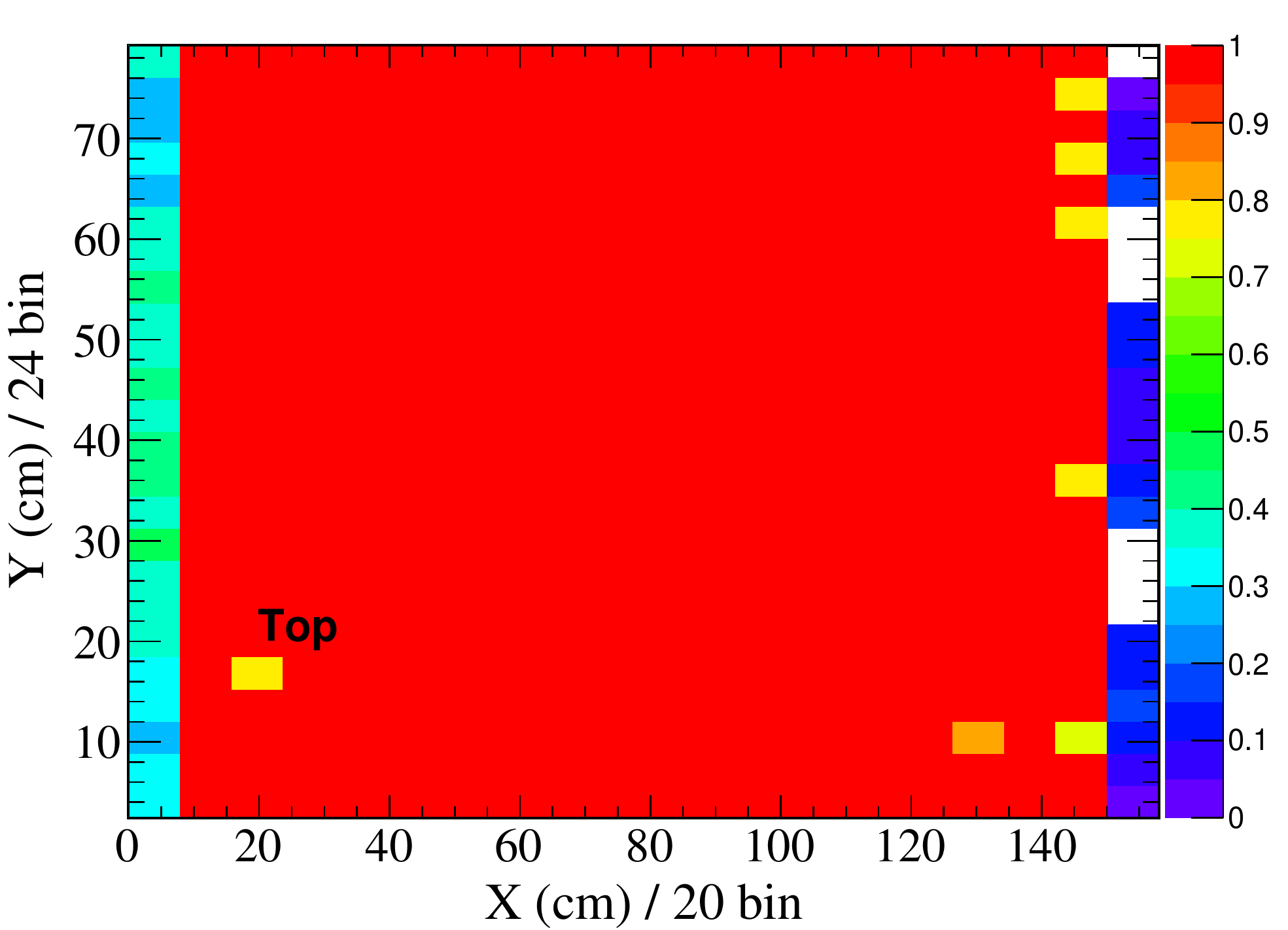}
\put(-100,90){Preliminary}
\includegraphics[width=0.33\columnwidth]{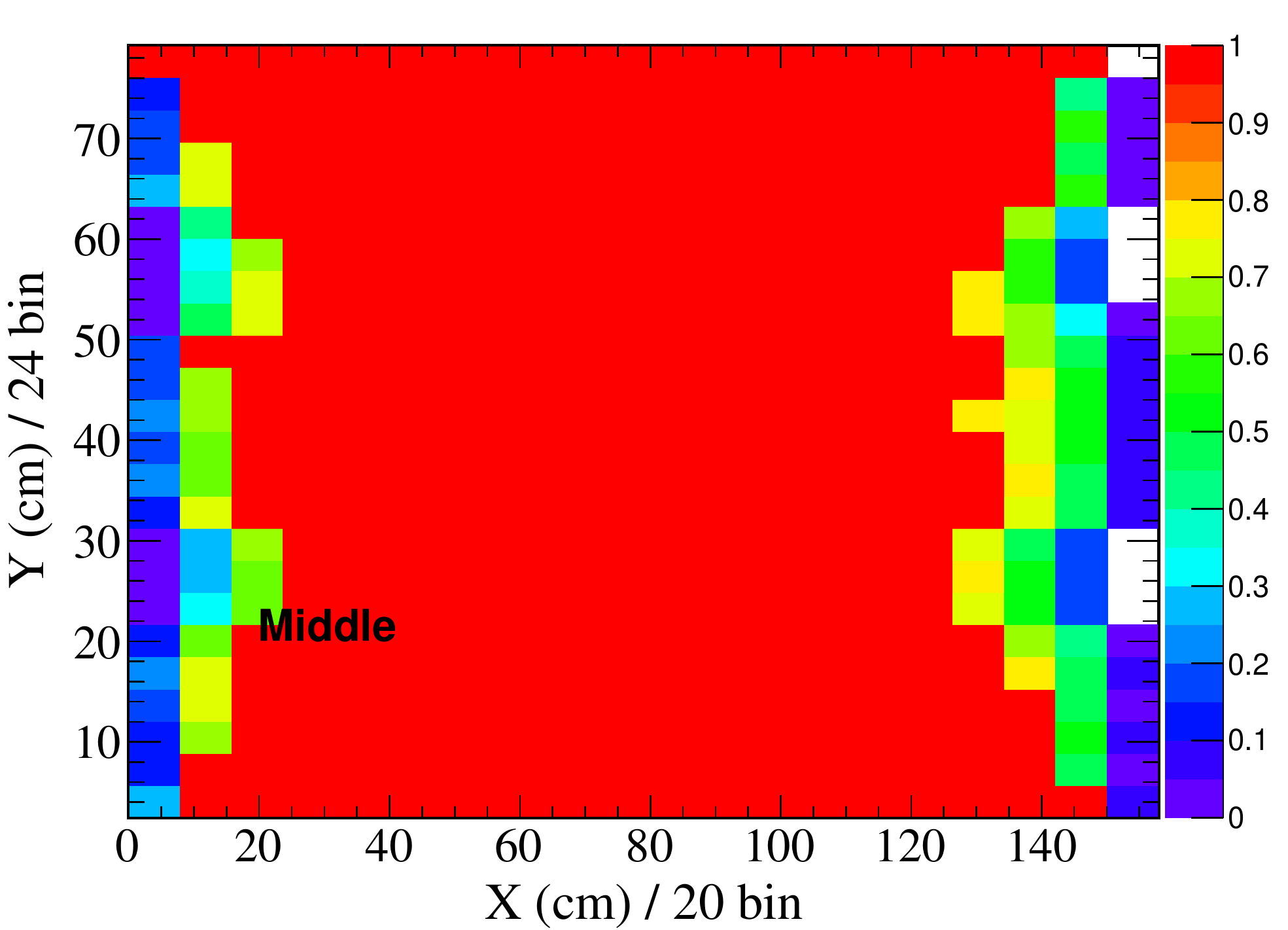}
\put(-100,90){Preliminary}
\includegraphics[width=0.33\columnwidth]{effB_Afftori3}
\put(-100,90){Preliminary}
\caption{Global efficiency map for top, middle and bottom chamber of TORI-03 telescope.} \label{effi}
\end{figure}

The global efficiency map, for each chamber, is  obtained as the product of the tracking and counting efficiency maps of the same chamber. The global efficiency of the three chambers of  the TORI-03 telescope are reported in figure \ref{effi} and these maps have been used to correct the polar angle distribution of simulated events. By looking the maps reported in figure \ref{effi}, it is evident as this procedure is able to identify the inefficiency at the edges of the middle chamber along the X direction, probably due to local gas leak, and the inefficiency of a strip in the bottom chamber.

\begin{figure}[h]
\centering
\includegraphics[width=0.5\columnwidth]{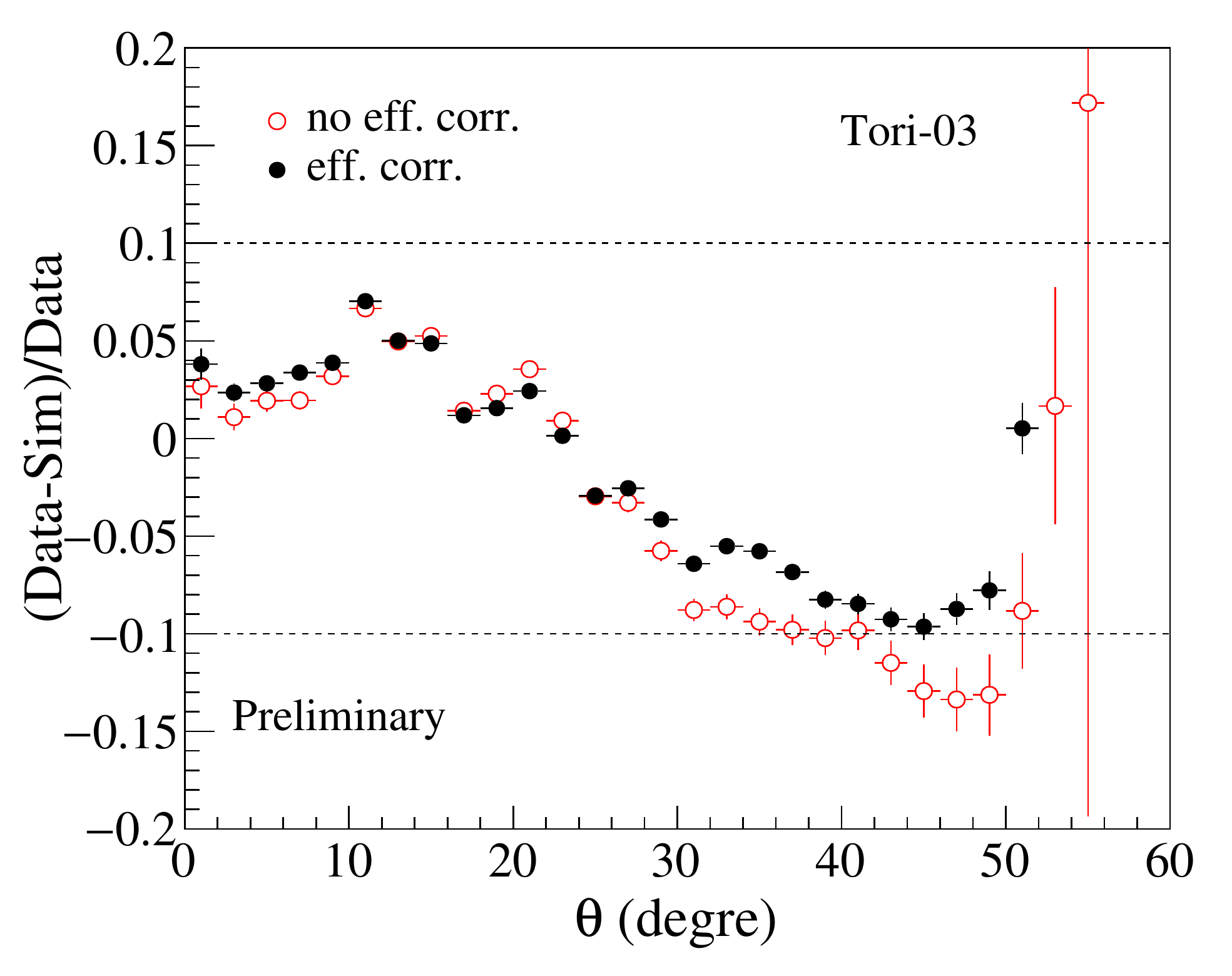}
\caption{Ratio between the experimental and simulated polar angle distributions with efficiency corrections (full circles) and without (open circles).} \label{efficorr}
\end{figure}

In figure \ref{efficorr}, we report the ratio between experimental and simulated polar angle distribution with and without efficiency correction. The efficiency corrections derived from data are able to improve the experimental-simulation agreement within 10\% in the whole polar angle acceptance of the  EEE telescope; further investigations are in progress to improve the description of our detector in the simulation tool.       
The improvement of the experimental-simulation agreement at large angle proves the reliability of the procedure to extract the efficiency maps of the chambers of the telescope by using the experimental data.

\section{Conclusion}
In this work we present a simulation tool based on GEMC developed to describe the MRPC telescope of the EEE experiment\cite{Ref1,Ref2}. This simulation code provides an event generation implemented by using an improved version of the Gaisser parametrization of cosmic muon flux as a function of muon energy and momentum\cite{egpar}. Moreover, this tool is able to describe the single telescope behaviour reproducing an important quantity such as the muon polar angle direction  with a precision of 10\% in the whole detector acceptance. 
A procedure to estimate the efficiency map of telescope derived directly from data has been presented and its reliability has been proved by comparing experimental and simulated data. The code is ready to be used with different event generators, like Corsika \cite{corsy}, in order to investigate the extensive cosmic rays showers.

\end{document}